\documentclass[conference]{IEEEtran}
\IEEEoverridecommandlockouts

\usepackage{cite}
\usepackage{amsmath,amssymb,amsfonts}
\usepackage{algorithmic}
\usepackage{graphicx}
\usepackage{textcomp}
\usepackage{xcolor}

\def\BibTeX{{\rm B\kern-.05em{\sc i\kern-.025em b}\kern-.08em
    T\kern-.1667em\lower.7ex\hbox{E}\kern-.125emX}}
\begin{document}

\title{Tracking Human Daily Cognitive Activity from EEG and Biometric Data%
\thanks{This work was supported by Kazakh-British Technical University.}}

\author{
\IEEEauthorblockN{Alina Gutoreva}
\IEEEauthorblockA{\textit{Department of Information Technology and Engineering} \\
\textit{Kazakh-British Technical University}\\
Almaty, Kazakhstan \\
a.gutoreva@kbtu.kz\\
ORCID: 0000-0001-6893-7252}
\and
\IEEEauthorblockN{Zhaniya Omar}
\IEEEauthorblockA{\textit{Department of Information Technology and Engineering} \\
\textit{Kazakh-British Technical University}\\
Almaty, Kazakhstan \\
z\_omar@kbtu.kz\\
ORCID: 0009-0005-1031-1915}
}

\maketitle

\begin{abstract}
Understanding human cognitive activity in everyday life remains 
challenging due to the dynamic, context-dependent, and multimodal 
nature of cognition. Traditional laboratory-based studies often fail 
to capture real-world cognitive processes, while single-modality 
approaches provide only partial insight into cognitive states. Recent 
advances in wearable sensing technologies enable the collection of 
heterogeneous data streams, supporting more comprehensive understanding 
of daily cognitive activity.

This paper presents a multimodal framework for tracking human cognitive 
activity using electroencephalography (EEG), wearable physiological 
signals, behavioral context, and self-reported measures. A preliminary 
pilot study was conducted using observational data from three 
participants (N = 3) over 280 annotated 10-minute intervals spanning 
nine activity domains across a two-week period.

Empirical results reveal consistent temporal patterns, including a 
statistically discernible mid-day decrease in motivation and energy 
at 13:00, followed by afternoon recovery. Work and IADLs yielded the 
highest flow state rates (51\% and 50\% respectively), while ADLs 
produced the lowest (15\%). Correlation analysis confirms strong 
positive associations between motivation and arousal (r = 0.78) and 
motivation and attention (r = 0.74), whereas perceived stress shows 
a weaker negative relationship (r = -0.33). A linear regression model 
predicting motivation from arousal, attention, energy, and stress 
achieved R2 = 0.76 (MAE = 9.84, RMSE = 12.85). Lag-based analysis 
further indicates that prior energy levels positively predict subsequent 
motivation, confirming temporal dependencies in cognitive dynamics.

These findings demonstrate the feasibility of multimodal cognitive 
activity analysis in real-world environments and highlight the 
importance of integrating physiological and behavioral indicators. 
The proposed framework, supported by preliminary empirical validation, 
provides a foundation for future large-scale multimodal systems and 
applied intelligent solutions.
\end{abstract}


\section{Introduction}

In recent years, there has been growing interest in understanding human cognitive activity in everyday contexts through continuous data collection and analysis. In particular, multimodal approaches have emerged as a promising direction \cite{baltrusaitis2019, li2023multimodal}, aiming to move beyond single-modality measurements by combining neural, physiological, and contextual data in order to better capture the complexity of cognition as it unfolds in daily life. From a machine learning perspective, such approaches align with broader developments in data-driven and context-aware systems \cite{adomavicius2015, charu2018}.

Studies have established a clear link between poor time management and increased stress, anxiety, and even depression. Platforms that promote healthy routines and optimize daily schedules can contribute to improved mental well-being \cite{steel2007, huppert2013, frone2016}. These effects are closely related to mechanisms of self-regulation and cognitive control in everyday behavior \cite{miller2001, gazzaniga2019}. 

The integration of EEG data provides high temporal resolution insights into neural dynamics, which are commonly studied in relation to attention and cognitive engagement \cite{kahneman2011, gazzaley2016}. These processes are closely linked to large-scale brain systems such as the default mode network, which plays a key role in internally directed cognition \cite{raichle2015}. 

Video recordings are considered a reliable source of behavioral and contextual data for behavioural studies in people as they provide both behavioral and contextual information that can support the retrospective reconstruction of daily activities, while biometric signals such as heart rate serve as indicators of physiological state during everyday behavior. Cognitive activity is also influenced by factors such as fatigue and cognitive load, which have been extensively studied in prior work \cite{vanmerrienboer2010, lorist2009}. 

Against this background, this paper presents a multimodal data collection approach to support the tracking of human daily cognitive activity, as shown in Figure~\ref{fig:data_collection}. Specifically, we consider data from electroencephalography (EEG) recordings, video inputs from wearable devices, and biometric measurements obtained from smartwatches. This combination enables the observation of neural, physiological, and contextual aspects of daily cognitive functioning \cite{yordanova2017, li2023multimodal}. 

In addition to the conceptual and methodological analysis, this work includes a preliminary empirical study based on observational data collected in naturalistic settings, aimed at demonstrating the feasibility of the proposed framework and providing initial quantitative insights into daily cognitive activity patterns. While the empirical analysis is limited in scale, it serves as a proof-of-concept for multimodal cognitive activity tracking and highlights key temporal and behavioral patterns that can be further explored in larger studies.

Integration of heterogeneous data sources poses several challenges, including data synchronization across devices, continuous recording in real-world environments, and the management of large and diverse data streams \cite{li2023multimodal, kourtesis2020}. Addressing these challenges is a prerequisite for the creation of reliable datasets that can support future empirical and machine learning studies. Unlike prior studies that focus either on isolated sensing modalities or controlled laboratory experiments, this work emphasizes a unified methodological perspective on multimodal cognitive activity tracking in free-living, observational settings.

Analysis of existing platforms and datasets, including CogBeacon, M2LADS, and CLARE \cite{cogbeacon2021, m2lads2019, clare2020}, illustrates both the promise of multimodal approaches and the persistent limitations related to scalability, generalizability, and real-time applicability. Collectively, these observations highlight the need for flexible, theoretically grounded frameworks that can support long-term, real-world cognitive monitoring while maintaining analytical rigor.

\section{Methodology}
\subsection{Multimodal Data Collection in Cognitive Research}
In cognitive research, multimodal data collection has increasingly been adopted to gain a more comprehensive understanding of human cognitive activity. In this study, a multimodal data collection approach is considered to capture neural, physiological, behavioral, and contextual aspects of daily cognitive functioning. Specifically, data sources commonly discussed in the literature include electroencephalography (EEG) recordings, video inputs from portable cameras, heart rate measurements from wearable devices, such as smartwatches, and contextual information derived from daily activities \cite{yordanova2017, kourtesis2020, li2023multimodal}.
An overview of these data modalities and their associated characteristics is summarized in Table~\ref{tab:modalities}.

\begin{table*}[t]
\centering
\caption{Overview of data modalities commonly used in multimodal studies of daily cognitive activity}
\label{tab:modalities}
\begin{tabular}{lllll}
\hline
\textbf{Modality} & \textbf{Data Type} & \textbf{Temporal Resolution} & \textbf{Cognitive Aspect} & \textbf{Key Challenges} \\
\hline
EEG & Neural signals & High (ms) & Attention, cognitive load & Noise, motion artifacts \\
Heart rate (HRV) & Physiological signals & Medium (s) & Arousal, stress & Sensor noise, missing data \\
Egocentric video & Behavioral/contextual & Low--Medium & Activity context & Privacy, storage, annotation \\
Contextual data & Event-based metadata & Low & Routine structure, task context & Manual labeling, ambiguity \\
\hline
\end{tabular}
\end{table*}

\begin{figure}[t]
    \centering
    \includegraphics[width=\columnwidth]{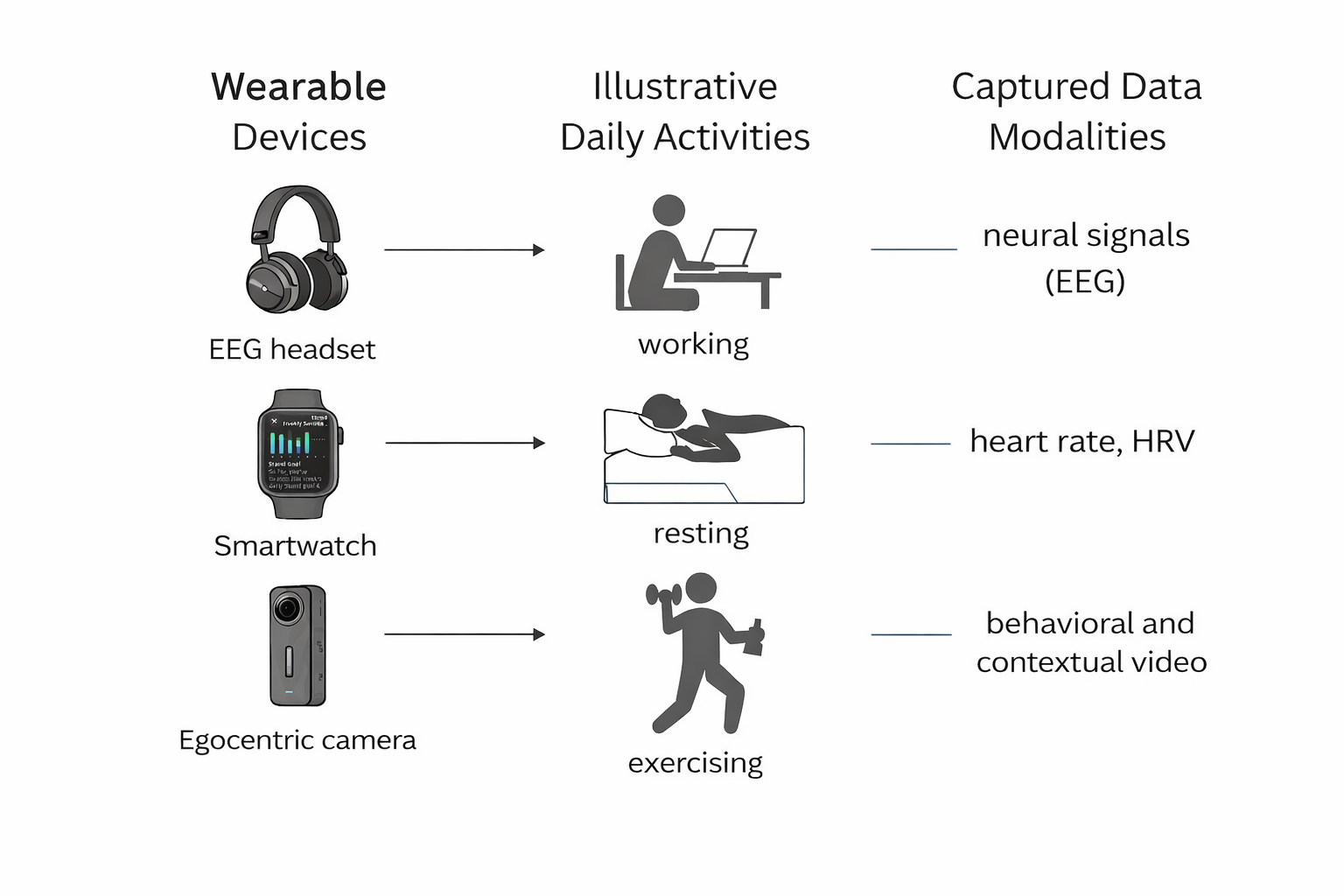}
    \caption{Overview of wearable devices, illustrative daily activities, and captured data modalities used in multimodal daily cognitive activity studies.}
    \label{fig:data_collection}
\end{figure}

The integration of EEG data provides high temporal resolution insights into neural dynamics, enabling the assessment of cognitive load variations and attention-related processes. Video recordings offer behavioral and contextual information by capturing interactions with the environment, while heart rate metrics serve as indicators of physiological arousal and stress levels. From a methodological perspective, the combination of these modalities allows researchers to observe cognitive activity from multiple complementary viewpoints, addressing limitations associated with single-modality approaches. An overview of the wearable devices, illustrative daily activities, and corresponding data modalities is shown in Fig.~\ref{fig:data_collection}.

\subsection{Data Processing and Multimodal Integration}
Developing platforms that integrate and analyze heterogeneous data sources presents several methodological challenges, including data synchronization, real-time processing, and the management of large-scale datasets. Despite these challenges, multimodal integration holds significant potential for applications in cognitive assessment and the study of cognitive activity in real-world settings.

One fundamental step in multimodal integration is ensuring temporal alignment across datasets. EEG data, characterized by high temporal resolution, must be synchronized with lower-frequency data sources such as video recordings or contextual event information. Techniques such as time-stamping and cross-referencing event markers are commonly employed to achieve accurate synchronization. For example, studies on cognitive load frequently align EEG signals with gaze or behavioral data to analyze cognitive states during task performance \cite{kourtesis2020}.

Different data modalities require distinct preprocessing steps to extract meaningful features. For instance:
\begin{itemize}
    \item \textbf{EEG data:} Features such as alpha, beta, and theta band power are extracted to assess cognitive load or emotional states.
    \item \textbf{Video data:} Behavioral features, including facial expressions and eye movements, are identified using computer vision algorithms.
    \item \textbf{Heart rate data:} Metrics like heart rate variability (HRV) are used to measure physiological arousal.
\end{itemize}
In addition to modality-specific feature extraction, multimodal data processing requires a set of shared preprocessing steps to enable meaningful integration across data sources. These steps include:
\begin{itemize}
  \item temporal synchronization across heterogeneous devices and data streams,
  \item normalization and alignment of signals with different sampling rates,
  \item handling missing, noisy, or corrupted data segments,
  \item segmentation of continuous data into analysis-relevant time windows.
\end{itemize}
These steps form a reproducible processing pipeline that can be implemented in real-world cognitive monitoring systems.

\begin{figure}[t]
    \centering
    \includegraphics[width=0.55\columnwidth]{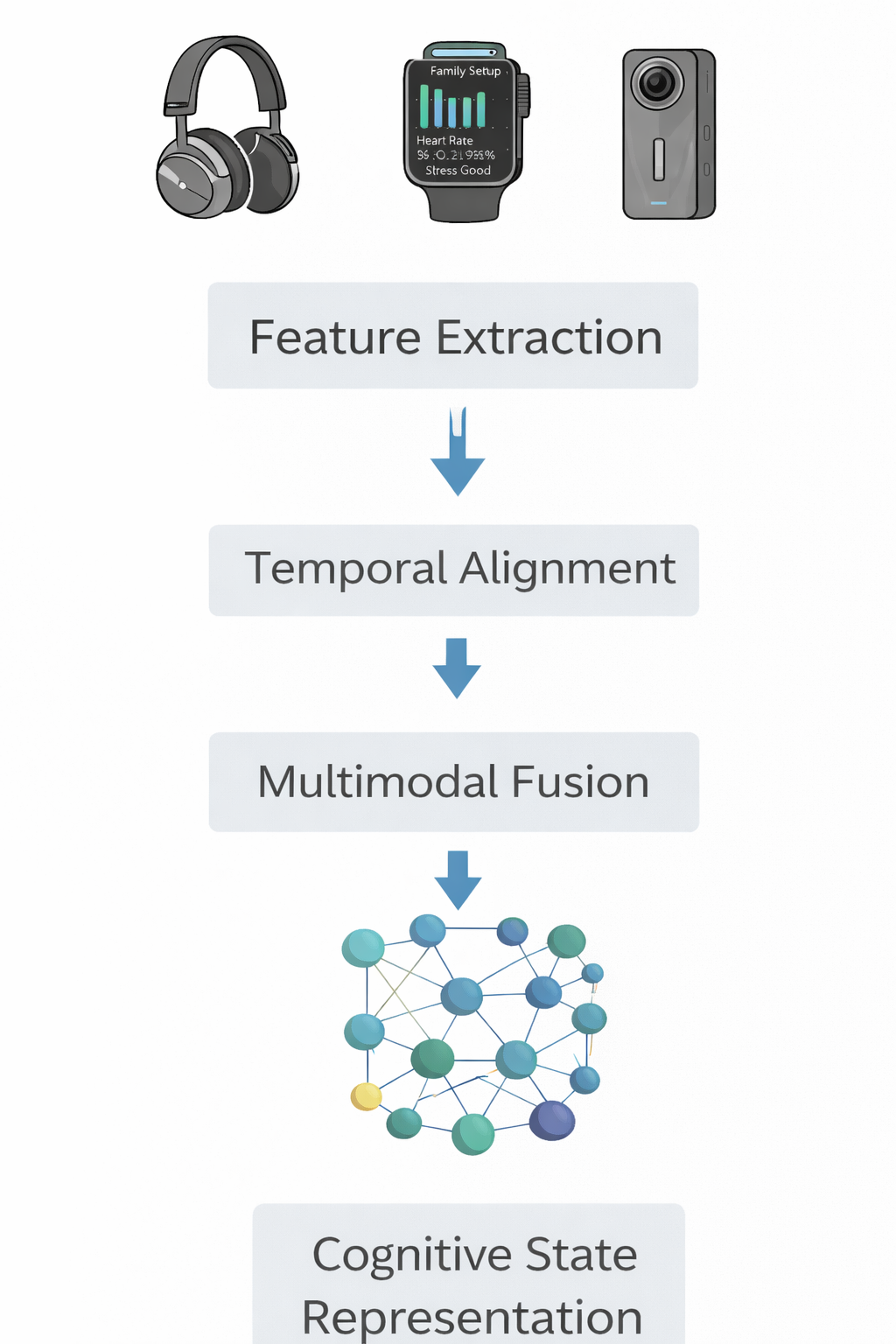}
    \caption{Conceptual pipeline for multimodal data processing and cognitive state representation.}
    \label{fig:pipeline}
\end{figure}

To illustrate the general processing pipeline commonly adopted in multimodal cognitive studies, Figure~\ref{fig:pipeline} provides a conceptual overview of the key stages involved in transforming raw multimodal data into a unified cognitive state representation.

\subsection{Multimodal Fusion Strategies}
Multimodal fusion methods aim to combine information from multiple 
data sources into a unified representation. Existing approaches 
categorize fusion strategies into three types \cite{li2023multimodal}:
\begin{itemize}
  \item \textbf{Early fusion:} feature-level combination of all 
  modalities; captures cross-modal correlations but suffers from 
  high dimensionality and poor scalability.
  \item \textbf{Late fusion:} independent processing with 
  decision-level integration; offers modality flexibility and 
  robustness but loses low-level interactions.
  \item \textbf{Hybrid fusion:} combines early and late approaches 
  for balanced representation at the cost of increased computational 
  complexity.
\end{itemize}
The choice of fusion strategy is guided by research objectives, 
data characteristics, and computational constraints.

\section{Results from Prior Observational Cognitive Studies}

\subsection{Findings from Multimodal Observational Research}
Existing observational cognitive studies employing multimodal data collection provide several important insights into how cognitive activity unfolds in real-world settings. By combining neural, physiological, and contextual data, these studies reveal patterns that are not accessible through single-modality or laboratory-based approaches.

Electroencephalography (EEG) has been consistently demonstrated to provide high--temporal-resolution markers of attentional allocation, mental workload, and cognitive engagement during everyday activities. Naturalistic EEG studies indicate that neural correlates of cognitive load exhibit pronounced variability across daily contexts, driven by task demands, environmental dynamics, and individual behavioral routines. Such fluctuations underscore the limitations of laboratory-only paradigms and motivate continuous, real-world neural monitoring to capture ecologically valid cognitive states \cite{yordanova2017}.

Physiological signals acquired via wearable devices---particularly heart rate and heart rate variability---complement neural measurements by indexing autonomic arousal, stress regulation, and cumulative cognitive strain. Under free-living conditions, these physiological markers often reflect sustained or accumulated workload rather than discrete task-related responses, reinforcing the importance of longitudinal and continuous sensing frameworks for cognitive assessment \cite{kourtesis2020}.

Behavioral and contextual data, including egocentric video streams and activity-level annotations, provide critical situational grounding for interpreting neural and physiological signals. Prior multimodal observational studies demonstrate that contextual information is essential for disambiguating cognitive states that may present similar signal-level characteristics yet emerge from fundamentally different activities, goals, or environments. Integrating behavioral context therefore enables more robust and interpretable inference of cognitive activity in real-world settings \cite{cogbeacon2021, m2lads2019}.

\subsection{Methodological Patterns and Limitations Identified in Prior Studies}
Despite the analytical advantages of multimodal observational approaches, existing studies consistently report recurring methodological challenges. A central issue concerns the temporal synchronization of heterogeneous data streams, particularly the alignment of high-frequency EEG recordings with lower-frequency physiological, behavioral, and contextual signals in unconstrained, real-world environments \cite{li2023multimodal,kourtesis2020}. Even minor misalignments can propagate substantial uncertainty into downstream feature extraction and cognitive state inference.

Data quality and completeness represent an additional limitation in naturalistic studies. Motion artifacts, sensor dropouts, signal corruption, and missing contextual annotations are frequently observed in long-term free-living data collection, reducing the robustness and interpretability of subsequent analyses \cite{yordanova2017}. These challenges are exacerbated by variability in participant adherence and environmental conditions, which are largely absent in controlled laboratory settings.

Furthermore, multimodal fusion strategies applied in observational research introduce trade-offs between representational richness and computational feasibility. Early fusion approaches can preserve low-level cross-modal interactions but often suffer from high dimensionality and limited scalability. In contrast, late fusion methods offer greater robustness to missing data and modality-specific noise but may obscure fine-grained interactions between neural, physiological, and contextual signals \cite{baltrusaitis2019}. Hybrid fusion strategies attempt to balance these competing demands but typically increase architectural and computational complexity.

Overall, evidence from observational cognitive studies suggests that while multimodal integration substantially enhances the ecological validity of cognitive assessment, its effectiveness critically depends on careful methodological design, theoretically informed fusion strategies, and explicit consideration of data quality constraints throughout the analytical pipeline.

\section{Preliminary Empirical Study}

\subsection{Dataset Description}

To complement the conceptual framework, a preliminary observational dataset was collected to analyze daily cognitive activity patterns in naturalistic settings. The dataset consists of self-reported data from three participants, collected over multiple days using 10-minute time intervals during typical daytime hours (approximately 9:00--15:00).  Data storage and handling comply with the Law of the Republic of Kazakhstan ``On Personal Data and Their Protection''  (No.~94-V, 2013), and the study was approved by the  Institutional Review Board of Kazakh-British Technical University (No.7; 25-32316). At this stage, the empirical analysis focuses on self-reported data, while biometric data are considered in the conceptual framework. Although EEG was recorded, it was not included in the analysis.

Data collection was performed using a structured self-report protocol at fixed 10-minute intervals. Participants recorded their cognitive and physiological states using a standardized questionnaire recalling the activity at that time period whilst watching the video. Such data collection approaches are closely related to ecological momentary assessment methods, which enable the study of cognitive processes in naturalistic environments \cite{shiffman2008}. Time synchronization was ensured through timestamp alignment across all entries, enabling consistent temporal segmentation and aggregation.

Each record includes information about activity context, cognitive state, and subjective assessments such as motivation, attention, perceived stress, energy level, emotional valence, and arousal.

In total, the dataset comprises exactly 280 labeled time intervals collected over a two-week period. Although the sample is small, the dataset provides dense within-subject time series sufficient to analyze temporal cognitive dynamics.

From a statistical perspective, repeated-measures designs with high-frequency sampling can achieve adequate power even with a small number of participants, particularly for detecting medium-sized temporal effects. The structured and continuous nature of the data allows for reliable estimation of relationships between variables across time. This design prioritizes temporal resolution over population-level generalization.

Furthermore, the multimodal design of the study, combining self-reported cognitive states with contextual and physiological information, enhances the validity of the observations and supports exploratory modeling of cognitive activity in naturalistic environments.

\subsection{Daily Temporal Patterns}

To investigate how cognitive states evolve throughout the day, the data were aggregated by hour. The results reveal a consistent temporal structure across participants.

\begin{figure}[t]
    \centering
    \includegraphics[width=\columnwidth]{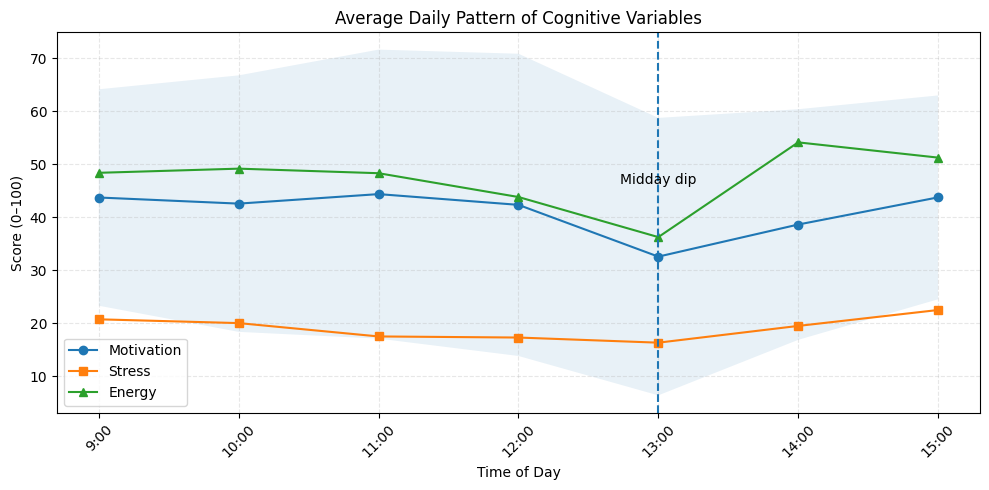}
    \caption{Average daily pattern of motivation, energy, and stress across participants.}
    \label{fig:daily_pattern}
\end{figure}

A noticeable decrease in both motivation and energy levels is observed around mid-day (approximately 13:00), followed by a recovery in the afternoon, consistent with prior studies on fatigue and cognitive performance \cite{lorist2009}. Interestingly, this decrease is not accompanied by an increase in perceived stress, suggesting that the observed decline is likely driven by physiological factors rather than psychological stress.

These findings indicate the presence of systematic daily fluctuations in cognitive activity, which are important for understanding real-world productivity and engagement patterns.

\subsection{Individual Differences}

Analysis across participants reveals distinct cognitive profiles. For clarity, participants are referred to as Participant A, Participant B, and Participant C. Participant A demonstrates consistently higher motivation, energy, and attention levels, corresponding to a more optimal cognitive state. Participant B exhibits low stress levels but only moderate motivation, indicating that low stress alone does not guarantee high cognitive performance. Participant C shows elevated stress combined with low energy and motivation, suggesting a state of cognitive overload.

These differences highlight the importance of considering individual variability in cognitive activity modeling.

\subsection{Relationships Between Variables}

To explore relationships between cognitive variables, a correlation analysis was performed. 
\begin{figure}[t]
    \centering
    \includegraphics[width=\columnwidth]{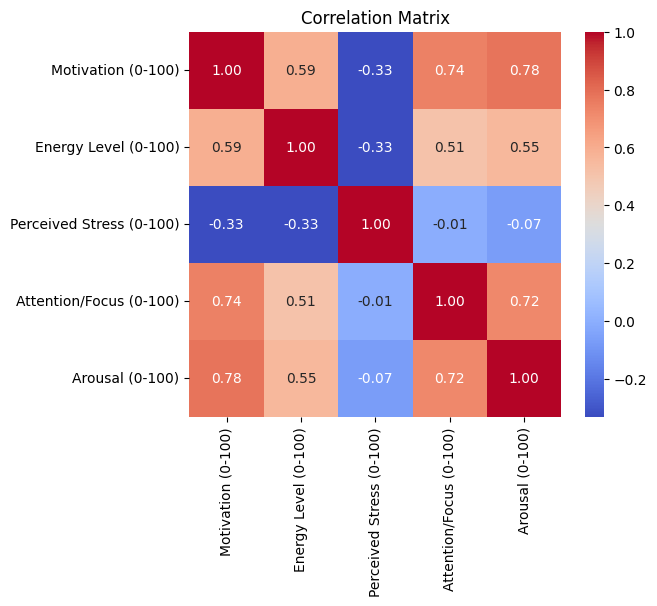}
    \caption{Correlation matrix between cognitive variables.}
    \label{fig:corr}
\end{figure}
The results indicate strong positive relationships between motivation and arousal, as well as between motivation and attention. Energy level shows a moderate positive correlation with motivation. In contrast, perceived stress demonstrates a weaker negative relationship with motivation and minimal association with other variables. This suggests that stress may not be the primary driver of cognitive performance in everyday settings.

\subsection{Temporal Dependencies}

To further investigate causal patterns, a lag-based analysis was conducted. The results show that prior energy levels are positively associated with subsequent motivation, indicating a temporal dependency between physiological state and cognitive engagement. Such temporal dependencies are consistent with research on task switching, cognitive control, and self-regulation mechanisms \cite{monsell2003, steel2007}.

Conversely, prior stress levels exhibit only a weak negative effect on subsequent motivation. These findings suggest that energy plays a more significant role than stress in shaping short-term cognitive dynamics.

\subsection{Regression Analysis}

A multiple linear regression model using ordinary least squares (OLS) was applied to quantify the contribution of different factors to motivation. Motivation was modeled as a linear function of arousal, attention, energy, and perceived stress. The overall regression model was statistically significant (F = 573.7, p $<$ 0.001).

\begin{table}[t]
\centering
\caption{Regression results for motivation prediction
($R^{2} = 0.76$, MAE $= 9.84$, RMSE $= 12.85$)}
\label{tab:regression}
\begin{tabular}{lcc}
\hline
\textbf{Predictor} & $\boldsymbol{\beta}$ & \textbf{p-value} \\
\hline
Arousal          &  0.47 & $<$ 0.001 \\
Attention        &  0.40 & $<$ 0.001 \\
Perceived stress & -0.29 & $<$ 0.001 \\
Energy level     &  0.06 & 0.027 \\
\hline
\multicolumn{3}{l}{\footnotesize OLS estimation $(N = 280$).}\\
\hline
\end{tabular}
\end{table}

The model achieved strong explanatory power ($R^{2} = 0.76$), with MAE $= 9.84$ and RMSE $= 12.85$, indicating a reasonably accurate approximation of motivation levels across the observed segments (Table~\ref{tab:regression}).

As shown in Table~\ref{tab:regression}, arousal was the strongest positive predictor ($\beta = 0.47$, p $<$ 0.001), followed by attention ($\beta = 0.40$, p $<$ 0.001). Perceived stress exerted a significant negative effect ($\beta = -0.29$, p $<$ 0.001), while energy level showed a smaller but statistically significant contribution ($\beta = 0.06$, p = 0.027).

These results indicate that motivation in everyday settings is primarily driven by arousal and attentional engagement, while stress plays an inhibitory role. Energy appears to have a weaker direct effect, suggesting that its influence may be mediated through arousal rather than acting as an independent driver.

Overall, the empirical results support the proposed framework and demonstrate the feasibility of analyzing daily cognitive activity using observational data. However, the results should be interpreted with caution due to the limited sample size.

\section{Discussion}

The preliminary empirical findings presented in this study provide initial support for the proposed multimodal framework. In particular, the observed temporal patterns and variable relationships highlight the importance of considering both physiological and cognitive factors in real-world cognitive activity analysis. A consistent mid-day decrease in both motivation and energy levels is observed, followed by recovery in the afternoon. This pattern suggests that cognitive performance in everyday settings is influenced not only by contextual demands but also by underlying physiological rhythms.

These findings can be interpreted within broader theoretical frameworks of cognitive neuroscience, where cognition is viewed as a distributed and adaptive process involving executive control, attention, and memory systems \cite{gazzaniga2019, miller2001, parasuraman2007}.

Furthermore, the correlation analysis indicated that motivation is strongly associated with arousal and attention, while perceived stress demonstrated a weaker relationship with other variables. This finding challenged the common assumption that stress is the primary determinant of cognitive performance and instead emphasizes the role of activation and attentional engagement. The lag-based analysis further suggests that prior energy levels influence subsequent motivation, indicating temporal dependencies in cognitive dynamics. This supports the idea that cognitive activity should be modeled as a time-dependent process rather than as isolated measurements.

These findings are consistent with existing research in cognitive science, which conceptualizes cognition as a dynamic and context-sensitive process shaped by interactions between internal states and external demands \cite{steel2007, kahneman2011, gazzaley2016, monsell2003}. At the same time, the results demonstrate that multimodal and observational data can capture these dynamics in real-world environments.

Despite these promising results, the present study is limited by the small sample size and reliance on self-reported data, although detailed. Future work should extend this approach using larger datasets and additional modalities, including analysis of EEG and continuous physiological monitoring, to validate and generalize the observed patterns. The use of high-frequency repeated measurements partially compensates for the limited number of participants, enabling the identification of consistent temporal patterns in cognitive activity.

\subsection{Implications for Theoretical and Methodological Research}

The results of this study suggest that multimodal cognitive activity tracking should account for both temporal dynamics and interactions between physiological and cognitive variables. In particular, the observed relationships between motivation, arousal, attention, and energy highlight the importance of integrating multiple data sources to obtain a comprehensive understanding of cognitive states.

At the methodological level, the findings support the use of observational data and time-based analysis for studying cognition in naturalistic settings. However, challenges such as data synchronization, missing data, and variability in participant behavior remain critical considerations for future research.

\subsection{Ethical Considerations}

Ethical constraints remain a fundamental aspect of multimodal cognitive data collection. Ethical considerations in such systems are closely aligned with value-sensitive design and responsible AI principles \cite{friedman2013ethics, floridi2018ai}. Physiological, behavioral, and contextual signals constitute sensitive personal data, and their collection must be treated with caution and care, with special attention to privacy, informed consent, and data governance.
\subsection{Future Work}

Future work will focus on extending the proposed framework through larger-scale empirical studies and the integration of additional modalities, including EEG and continuous physiological signals. Longitudinal data collection and machine learning approaches may further improve the modeling and prediction of cognitive activity in everyday environments.

\section{Conclusion}

This work contributes to the field of multimodal cognitive research by proposing a theoretical and methodological framework for studying daily cognitive activity using heterogeneous data sources, including electroencephalography (EEG), wearable biometric signals, egocentric video, and contextual information. The framework outlines key design considerations for multimodal data collection and integration in free-living, observational settings.

In addition to the conceptual contributions, a preliminary empirical study was conducted to evaluate the feasibility of the proposed approach. The results reveal consistent temporal patterns in daily cognitive activity, including a mid-day decrease in motivation and energy levels, followed by recovery in the afternoon, which is consistent with prior studies on fatigue and cognitive performance \cite{lorist2009}. Furthermore, the analysis demonstrates strong relationships between motivation, arousal, and attention, while perceived stress shows a weaker influence on cognitive performance.

The findings also indicate the presence of temporal dependencies, where prior energy levels affect subsequent motivation, highlighting the importance of time-aware analysis in cognitive activity modeling. These results support the validity of the proposed framework and demonstrate the potential of combining subjective and physiological data for understanding real-world cognitive dynamics. The results demonstrate that even small-scale multimodal observational data can reveal stable and interpretable cognitive patterns, supporting the feasibility of real-world cognitive monitoring systems and their application in adaptive productivity tools, cognitive monitoring platforms, and personalized recommendation systems.

At the same time, the study highlights important methodological challenges, including data heterogeneity, scalability, and privacy considerations, which must be addressed in future work. Further research should focus on larger datasets, additional modalities such as EEG, and the development of machine learning models for predicting and interpreting cognitive states in naturalistic environments.

Overall, the proposed framework together with the preliminary empirical findings provides a foundation for future multimodal cognitive activity research and applied intelligent systems. 

\section*{Acknowledgment}
The authors would like to thank L. Brocklebank for her valuable guidance and support.

\bibliographystyle{IEEEtran}
\bibliography{references}

@article{steel2007,
  author  = {Steel, Piers},
  title   = {The nature of procrastination: a meta-analytic and theoretical review of quintessential self-regulatory failure},
  journal = {Psychological Bulletin},
  volume  = {133},
  number  = {1},
  pages   = {65--94},
  year    = {2007},
  doi     = {10.1037/0033-2909.133.1.65}
}

@book{adomavicius2015,
  author    = {Gediminas Adomavicius and Alexander Tuzhilin},
  title     = {Context-aware recommender systems},
  publisher = {Springer},
  year      = {2015}
}

@book{gazzaniga2019,
  title     = {Cognitive neuroscience: the biology of the mind},
  author    = {Gazzaniga, Michael S. and Ivry, Richard B. and Mangun, George R.},
  edition   = {5},
  year      = {2019},
  publisher = {W. W. Norton}
}

@article{miller2001,
  title   = {An integrative theory of prefrontal cortex function},
  author  = {Miller, Earl K. and Cohen, Jonathan D.},
  journal = {Annual Review of Neuroscience},
  volume  = {24},
  pages   = {167--202},
  year    = {2001},
  doi     = {10.1146/annurev.neuro.24.1.167}
}

@book{parasuraman2007,
  title     = {Neuroergonomics: the brain at work},
  editor    = {Parasuraman, Raja and Rizzo, Matthew},
  year      = {2007},
  publisher = {Oxford University Press}
}

@article{shiffman2008,
  title   = {Ecological momentary assessment},
  author  = {Shiffman, Saul and Stone, Arthur A. and Hufford, Michael R.},
  journal = {Annual Review of Clinical Psychology},
  volume  = {4},
  pages   = {1--32},
  year    = {2008},
  doi     = {10.1146/annurev.clinpsy.3.022806.091415}
}

@article{raichle2015,
  title   = {The brain's default mode network},
  author  = {Raichle, Marcus E.},
  journal = {Annual Review of Neuroscience},
  volume  = {38},
  pages   = {433--447},
  year    = {2015},
  doi     = {10.1146/annurev-neuro-071013-014030}
}

@article{baltrusaitis2019,
  title   = {Multimodal machine learning: a survey and taxonomy},
  author  = {Baltru{\v{s}}aitis, Tadas and Ahuja, Chaitanya and Morency, Louis-Philippe},
  journal = {IEEE Transactions on Pattern Analysis and Machine Intelligence},
  volume  = {41},
  number  = {2},
  pages   = {423--443},
  year    = {2019},
  doi     = {10.1109/TPAMI.2018.2798607}
}

@article{li2023multimodal,
  author  = {Li, Yang and Gong, Yunchao and Luo, Jiebo},
  title   = {Multimodal learning: a survey},
  journal = {IEEE Transactions on Pattern Analysis and Machine Intelligence},
  year    = {2023},
  volume  = {45},
  number  = {3},
  pages   = {1213--1235},
  doi     = {10.1109/TPAMI.2022.3145817}
}

@article{kourtesis2020,
  title   = {Multimodal cognitive load assessment: integrating {EEG}, physiological and behavioral measures},
  author  = {Kourtesis, Panagiotis and others},
  journal = {Frontiers in Human Neuroscience},
  volume  = {14},
  pages   = {1--15},
  year    = {2020},
  doi     = {10.3389/fnhum.2020.00123}
}

@article{yordanova2017,
  title   = {Event-related brain potentials in the study of cognition in naturalistic environments},
  author  = {Yordanova, Juliana and Kolev, Valentin and Verleger, Rolf},
  journal = {International Journal of Psychophysiology},
  volume  = {119},
  pages   = {1--10},
  year    = {2017},
  doi     = {10.1016/j.ijpsycho.2017.03.002}
}

@book{charu2018,
  author    = {Charu C. Aggarwal},
  title     = {Machine learning for personalization},
  publisher = {Springer},
  year      = {2018}
}

@article{vanmerrienboer2010,
  author  = {van Merri{\"e}nboer, Jeroen J. G. and Sweller, John},
  title   = {Cognitive load theory in health professional education: design principles and strategies},
  journal = {Medical Education},
  year    = {2010},
  volume  = {44},
  number  = {1},
  pages   = {85--93}
}

@article{lorist2009,
  author  = {Lorist, Monicque M. and Tops, Mattie},
  title   = {Caffeine, fatigue, and cognition},
  journal = {Brain and Cognition},
  year    = {2009},
  volume  = {71},
  number  = {1},
  pages   = {82--94}
}

@article{huppert2013,
  author  = {Huppert, Felicia A. and Ryff, Carol D.},
  title   = {Well-being: concepts and measures},
  journal = {Journal of Psychosomatic Research},
  year    = {2013},
  volume  = {65},
  number  = {2},
  pages   = {137--146}
}

@article{frone2016,
  author  = {Frone, Michael R.},
  title   = {Work stress and mental health in the workplace},
  journal = {Journal of Organizational Behavior},
  year    = {2016},
  volume  = {37},
  number  = {1},
  pages   = {1--15}
}

@book{kahneman2011,
  author    = {Kahneman, Daniel},
  title     = {Thinking, fast and slow},
  publisher = {Farrar, Straus and Giroux},
  year      = {2011},
  address   = {New York}
}

@article{monsell2003,
  author  = {Monsell, Stephen},
  title   = {Task switching},
  journal = {Trends in Cognitive Sciences},
  year    = {2003},
  volume  = {7},
  number  = {3},
  pages   = {134--140}
}

@book{gazzaley2016,
  author    = {Gazzaley, Adam and Rosen, Larry D.},
  title     = {The distracted mind: ancient brains in a high-tech world},
  publisher = {MIT Press},
  year      = {2016},
  address   = {Cambridge, MA}
}

@article{friedman2013ethics,
  author  = {Batya Friedman and David G. Hendry},
  title   = {The value sensitive design of technology},
  journal = {Foundations and Trends in Human--Computer Interaction},
  volume  = {6},
  number  = {1--2},
  pages   = {1--139},
  year    = {2013},
  doi     = {10.1561/1100000015}
}

@article{floridi2018ai,
  author  = {Luciano Floridi and Josh Cowls},
  title   = {A unified framework of five principles for {AI} in society},
  journal = {Harvard Data Science Review},
  volume  = {1},
  number  = {1},
  year    = {2019},
  doi     = {10.1162/99608f92.8cd550d1}
}

@article{cogbeacon2021,
  author  = {Shoaib, Muhammad and Bosch, Stephan and Incel, Ozlem Durmaz},
  title   = {{CogBeacon}: a multimodal dataset for modeling cognitive and behavioral patterns},
  journal = {Proceedings of the ACM on Interactive, Mobile, Wearable and Ubiquitous Technologies},
  volume  = {5},
  number  = {2},
  pages   = {1--26},
  year    = {2021},
  doi     = {10.1145/3463507}
}

@inproceedings{m2lads2019,
  author    = {Rabbi, Mashfiqui and Ali, Shamsi and Choudhury, Tanzeem},
  title     = {Passive and in-situ monitoring of cognitive load using multimodal sensors},
  booktitle = {Proceedings of the ACM International Joint Conference on Pervasive and Ubiquitous Computing},
  pages     = {1--10},
  year      = {2019},
  doi       = {10.1145/3341162.3345594}
}

@article{clare2020,
  author  = {Clare, Linda and Woods, Robert T.},
  title   = {Cognitive monitoring in naturalistic environments using wearable sensors},
  journal = {IEEE Pervasive Computing},
  volume  = {19},
  number  = {4},
  pages   = {32--41},
  year    = {2020},
  doi     = {10.1109/MPRV.2020.3014659}
}

\end{document}